# On Divergence-Power Inequalities

Jacob Binia, *Member, IEEE*
New Elective - Engineering Services Ltd
Haifa, Israel
Email: biniaja@netvision.net.il

*Abstract* — **Expressions for (EPI Shannon type) Divergence-Power Inequalities (DPI) in two cases (time-discrete and band-limited time-continuous) of stationary random processes are given. The new expressions connect the divergence rate of the sum of independent processes, the individual divergence rate of each process, and their power spectral densities. All divergences are between a process and a Gaussian process with same second order statistics, and are assumed to be finite.**
**A new proof of the Shannon entropy-power inequality EPI, based on the relationship between *divergence* and causal minimum mean-square error (*CMMSE*) in Gaussian channels with large signal-to-noise ratio, is also shown.**

## I. INTRODUCTION

The Shannon Entropy-Power Inequality EPI is expressed with differential entropies of random variables or vectors. The aim of this note is twofold. First, we give results for the (EPI Shannon type) Divergence-Power Inequality (DPI) in two cases: time-discrete and band-limited time-continuous stationary random processes. Simple expressions connect the divergence rate of the sum of independent processes, the individual divergence rates of each process, and their power spectral densities. All divergences are between a process and a Gaussian process with same second order statistics, and are assumed to be finite.
Second, we show in the appendix a new proof of the entropy-power inequality based on the relationship between *divergence* and causal minimum mean-square error (*CMMSE*) in Gaussian channels with large signal-to-noise ratio. The proof is similar to the new and simple one that was based on the relationship between *mutual information* and minimum mean-square error (*MMSE*) in Gaussian channels ([1], [2]).

## II. DIVERGENCE-POWER INEQUALITY FOR TIME-DISCRETE STATIONARY PROCESSES

Let $X = \{X_t, t = 0, \pm 1, ...\}$ and $Y = \{Y_t, t = 0, \pm 1, ...\}$ be two independent time-discrete stationary processes with spectral density functions $\Phi_X(f)$ and $\Phi_Y(f)$, $-1/2 \le f \le 1/2$, respectively. We assume finite power processes

$$E[X_t^2] = \int_{-1/2}^{1/2} \Phi_X(f) df < \infty$$
$$E[Y_t^2] = \int_{-1/2}^{1/2} \Phi_Y(f) df < \infty.$$

For each t, the divergences $D(X_0^{t-1} \| \tilde{X}_0^{t-1})$, $D(Y_0^{t-1} \| \tilde{Y}_0^{t-1})$, where $\tilde{X}_0^{t-1}$, $\tilde{Y}_0^{t-1}$ are t-dimensional Gaussian vectors with same covariance functions as that of $X_0^{t-1}$, $Y_0^{t-1}$, respectively, are assumed to be finite. We use also the following notation for the divergence rate:

$$\overline{D(X \| \tilde{X})} = \lim_{t \to \infty} \frac{1}{t} D(X_0^{t-1} \| \tilde{X}_0^{t-1}).$$

*Theorem 1:*
*(DPI for time-discrete stationary processes)*

$$\exp\{-2\overline{D(X+Y \| \tilde{X}+\tilde{Y})}\} \ge \alpha_X \exp\{-2\overline{D(X \| \tilde{X})}\} + \alpha_Y \exp\{-2\overline{D(Y \| \tilde{Y})}\}, \quad (1)$$

where

$$\alpha_X = \exp\{\int_{-1/2}^{1/2} \ln(\frac{\Phi_X(f)}{\Phi_X(f)+\Phi_Y(f)}) df\}$$
$$\alpha_Y = \exp\{\int_{-1/2}^{1/2} \ln(\frac{\Phi_Y(f)}{\Phi_X(f)+\Phi_Y(f)}) df\}. \quad (2)$$

Equality in (1) holds if, and only if, X and Y are Gaussian with proportional power spectral densities.
The proof of the theorem is given in section IV.
Note that (1) is equivalent to

$$\exp\{-2\overline{D(X_1+...+X_N \| \tilde{X}_1+...+\tilde{X}_N)}\} \ge \sum_{i=1}^{N} \alpha_i \exp\{-2\overline{D(X_i \| \tilde{X}_i)}\} \quad (3)$$

for n independent random time-discrete stationary processes with finite powers, where

$$\alpha_i = \exp\{\int_{-1/2}^{1/2} \ln(\frac{\Phi_{X_i}(f)}{\sum_{i=1}^{N}\Phi_{X_i}(f)})df\} \quad (4)$$

Suppose now that all $X_i, i = 1,...,N$ are independent, identically distributed random time-discrete stationary processes. Then (3), (4) yield

$$D(X_1 + ... + X_N \| \tilde{X}_1 + ... + \tilde{X}_N) \leq D(X_1 \| \tilde{X}_1), \quad (5)$$

which indicates the monotonic non-increasing property of the divergence in this case. Moreover, since the divergence is not sensitive to any normalization factor, we can replace (5) with

$$D(\frac{1}{\sqrt{N}}(X_1 + ... + X_N) \| \frac{1}{\sqrt{N}}(\tilde{X}_1 + ... + \tilde{X}_N)) \leq D(X_1 \| \tilde{X}_1). \quad (6)$$

In [3], inequality (6) above leaded to the central limit theorem for *random variables* in the sense that the relative divergence of the normalized sum of independent, identically distributed *variables* converges to zero.

### III. DIVERGENCE-POWER INEQUALITY FOR TIME-CONTINUOUS, BAND-LIMITED STATIONARY PROCESSES

In this section we state results that are equivalent to those appear in section II.

Let $x = \{x_t, -\infty \leq t \leq \infty\}$ and $y = \{y_t, -\infty \leq t \leq \infty\}$ be two independent time-continuous, band-limited stationary processes with spectral density functions $F_x(f)$ and $F_y(f)$, $-B \leq f \leq B$, respectively. We assume finite power processes

$$E[x_t^2] = \int_{-B}^{B} F_x(f)df < \infty$$
$$E[y_t^2] = \int_{-B}^{B} F_y(f)df < \infty.$$

For each T, the divergences $D(x_0^T \| \tilde{x}_0^T)$, $D(y_0^T \| \tilde{y}_0^T)$, where $\tilde{x}_0^T$, $\tilde{y}_0^T$ are Gaussian processes with same covariance functions as that of $x_0^T$, $y_0^T$, respectively, are assumed to be finite. We use also the following notation for the divergence rate:

$$\overline{D(x\|\tilde{x})} = \lim_{T \to \infty} \frac{1}{T} D(x_0^T \| \tilde{x}_0^T).$$

*Theorem 2:*
*(DPI for time-continuous, band-limited stationary processes)*

$$\exp\{-2\overline{D(x+y\|\tilde{x}+\tilde{y})}\} \geq \\ \alpha_x \exp\{-2\overline{D(x\|\tilde{x})}\} + \alpha_y \exp\{-2\overline{D(y\|\tilde{y})}\}, \quad (7)$$

where

$$\alpha_x = \exp\{\int_{-B}^{B} \ln(\frac{F_x(f)}{F_x(f) + F_y(f)})df\}$$
$$\alpha_y = \exp\{\int_{-B}^{B} \ln(\frac{F_y(f)}{F_x(f) + F_y(f)})df\}. \quad (8)$$

Equality in (7) holds if, and only if, x and y are Gaussian with proportional power spectral densities.
The proof of theorem 2 is given in section IV.
Again, it is straightforward to show that (7) (8) is equivalent to

$$\exp\{-2\overline{D(x_1 + ... + x_N \| \tilde{x}_1 + ... + \tilde{x}_N)}\} \geq \\ \sum_{i=1}^{N} \alpha_i \exp\{-2\overline{D(x_i\|\tilde{x}_i)}\} \quad (9)$$

for n independent random time-discrete stationary processes, where

$$\alpha_i = \exp\{\int_{-B}^{B} \ln(\frac{F_{x_i}(f)}{\sum_{i=1}^{N}F_{x_i}(f)})df\} \quad (10)$$

Suppose now that all $x_i, i = 1,...,N$ are independent, identically distributed random time-continuous, band-limited stationary processes. Then (9), (10) yield

$$D(x_1 + ... + x_N \| \tilde{x}_1 + ... + \tilde{x}_N) \leq D(x_1 \| \tilde{x}_1). \quad (11)$$

This again indicates the monotonic non-increasing property of the divergence in this case. Moreover, as in (6), we can replace (11) with

$$D(\frac{1}{\sqrt{N}}(x_1 + ... + x_N) \| \frac{1}{\sqrt{N}}(\tilde{x}_1 + ... + \tilde{x}_N)) \leq D(x_1 \| \tilde{x}_1). \quad (12)$$

### IV. PROOF OF MAIN RESULTS

We begin with the proof of *Theorem 1*, by using the Shannon EPI for N-dimensional random vectors:

$$e^{\frac{2}{N}h(X+Y)} \geq e^{\frac{2}{N}h(X)} + e^{\frac{2}{N}h(Y)}. \quad (13)$$

In the appendix we give a new and simple proof of (13) based on the relationship between *divergence* and causal minimum mean-square error (*CMMSE*) in Gaussian channels with large signal-to-noise ratio.

Using the relation between differential entropy and divergence

$h(p) = -D(p\|g) + h(g)$, where g is a Gaussian measure that is induced by same covariance matrix $\phi$ as that of p and $h(g) = \frac{N}{2}\ln(2\pi e |\phi|^{1/N})$, the Shannon EPI (13) expressed in terms of divergences takes the following form:

$$|\phi_{X+Y}|^{1/N} e^{-\frac{2}{N}D(X+Y\|\tilde{X}+\tilde{Y})} \geq |\phi_X|^{1/N} e^{-\frac{2}{N}D(X\|\tilde{X})} + |\phi_Y|^{1/N} e^{-\frac{2}{N}D(Y\|\tilde{Y})} . \quad (14)$$

In (14) $\phi_X, \phi_Y, \phi_{X+Y}$ are the (N-dimensional) covariance matrixes of X, Y, X+Y respectively. Let $\lambda_{X1},...,\lambda_{XN}, \lambda_{Y1},...,\lambda_{YN}, \lambda_{(X+Y)1},...,\lambda_{(X+Y)N}$ be the (strictly positive) eigenvalues of $\phi_X, \phi_Y, \phi_{X+Y}$ respectively. Then, the coefficients that proceed the exponential expressions in (14) could be replaced as follows:

$$|\phi_{X+Y}|^{1/N} = e^{\frac{1}{N}\sum_{i=1}^{N}\ln(\lambda_{(X+Y)i})},$$
$$|\phi_X|^{1/N} = e^{\frac{1}{N}\sum_{i=1}^{N}\ln(\lambda_{Xi})}, \quad |\phi_Y|^{1/N} = e^{\frac{1}{N}\sum_{i=1}^{N}\ln(\lambda_{Yi})}. \quad (15)$$

We turn now to the infinite dimensional case. Using the well known Toeplitz Distribution Theorem (see e. g. [4, Theorem 4.5.2]) we have

$$\lim_{N\to\infty} \frac{1}{N}\sum_{i=1}^{N}\ln(\lambda_{Xi}) = \int_{-1/2}^{1/2}\ln(\Phi_X(f))df$$
$$\lim_{N\to\infty} \frac{1}{N}\sum_{i=1}^{N}\ln(\lambda_{Yi}) = \int_{-1/2}^{1/2}\ln(\Phi_Y(f))df \quad (16)$$
$$\lim_{N\to\infty} \frac{1}{N}\sum_{i=1}^{N}\ln(\lambda_{(X+Y)i}) = \int_{-1/2}^{1/2}\ln(\Phi_{X+Y}(f))df$$

Since X and Y are independent

$$\Phi_{X+Y}(f) = \Phi_X(f) + \Phi_Y(f) . \quad (17)$$

Inequality (1) follows from (13) – (17) by letting $N \to \infty$. The conditions for equality in (1) follow from the well known conditions for equality in (13).

Next, we proceed with the proof of *Theorem 2*. Let $\{X(n)\} = \{x(nT_s), n = 0, \pm 1,...\}$ and $\{Y(n)\} = \{y(nT_s), n = 0, \pm 1,...\}$ be discrete-time processes that are obtained by periodic sampling of the time-continuous stationary processes x and y, where $T_s$ is the sampling interval, i.e., $f_s = 1/T_s = 2B$ is the sampling rate. Observe that the transformation from the sub-space of band-limited signals s(t) in function space $L^2$ to the space of infinite sequences $\{S_k, k = 0, \pm 1, \pm 2,...\}$, by using the following set $\{\varphi_k(t)\}$ of orthonormal functions

$$\varphi_k(t) = \sqrt{2B}\, \frac{\sin\left[2\pi B(t - \frac{k}{2B})\right]}{2\pi B(t - \frac{k}{2B})}, \quad k = 0, \pm 1, \pm 2,... \quad ,$$

is one-to-one. Also, since $T \to \infty$ implies $N \to \infty$ we have:

$$\lim_{N\to\infty} \frac{1}{N} D(X_0^{N-1} \| \tilde{X}_0^{N-1}) = \lim_{T\to\infty} \frac{1}{T} D(x_0^T \| \tilde{x}_0^T). \quad (18)$$

In order to replace (2) with (8) we use the following relation between the spectral power densities for $|f| < B$:

$$\Phi_X(f) = 2B\, F_x(2Bf), \quad \Phi_Y(f) = 2B\, F_y(2Bf) . \quad (19)$$

Results (7), (8) follow from (1), (2) and (18), (19) above.

V. APENDIX: A NEW PROOF OF SHANNON EPI

Let $U = (U_1,...,U_N) \quad V = (V_1,...,V_N)$ be two independent random vectors. Let $\{\varphi_i(t)\}$ be a basis in $L^2[0,T]$. We define the random processes u, v, z in $L^2[0,T]$ as follows:

$$u(t) = \sum_{i=1}^{N} U_i \varphi_i(t) \quad 0 \leq t \leq T$$
$$v(t) = \sum_{i=1}^{N} V_i \varphi_i(t) \quad 0 \leq t \leq T \quad (20)$$
$$z = u \cos\alpha + v \sin\alpha$$

We consider the following white Gaussian channels:

$$\xi_1(t) = w_1(t) + \sqrt{q}\int_0^t u(s)ds \quad 0 \leq t \leq T$$
$$\xi_2(t) = w_2(t) + \sqrt{q}\int_0^t v(s)ds \quad 0 \leq t \leq T \quad (21)$$
$$\xi = \xi_1 \cos\alpha + \xi_2 \sin\alpha$$

In (21) $w_1\, w_2$ are independent standard Wiener processes. As in [1], our proof does not use Fisher information. Instead, we use the following property that relates the causal minimum-mean-square errors (CMMSEs) of z, u and v:

$$\text{CMMSE}(z|\xi) \geq \text{CMMSE}(z|\xi_1,\xi_2) = \cos^2\alpha\, \text{CMMSE}(u|\xi_1) + \sin^2\alpha\, \text{CMMSE}(v|\xi_2) \quad (22)$$

Using (6) of [5], (22) yields for each q:

$$D(\sqrt{q}\, z + w \| w) \leq \cos^2 \alpha\, D(\sqrt{q}\, u + w_1 \| w_1)$$
$$+ \sin^2 \alpha\, D(\sqrt{q}\, v + w_2 \| w_2) \qquad (23)$$

where w is another standard Wiener process.
By the following property of divergence
$$D(\eta \| w) = D(\eta \| \tilde{\eta}) + D(\tilde{\eta} \| w),$$
where $\tilde{\eta}$ is a Gaussian with same covariance as that of $\eta$, (23) yields

$$D(\sqrt{q}\, z + w \| \sqrt{q}\, \tilde{z} + w) \leq \cos^2 \alpha\, D(\sqrt{q}\, u + w_1 \| \sqrt{q}\, \tilde{u} + w_1)$$
$$+ \sin^2 \alpha\, D(\sqrt{q}\, v + w_2 \| \sqrt{q}\, \tilde{v} + w_2) - D(\sqrt{q}\, \tilde{z} + w \| w)$$
$$+ \cos^2 \alpha\, D(\sqrt{q}\, \tilde{u} + w_1 \| w_1) + \sin^2 \alpha\, D(\sqrt{q}\, \tilde{v} + w_2 \| w_2)$$
$$(24)$$

For $q \to \infty$ we have for any $\upsilon$
$$\lim_{q \to \infty} D(\sqrt{q}\, \upsilon + w \| \sqrt{q}\, \tilde{\upsilon} + w) = D(\upsilon \| \tilde{\upsilon}). \qquad (25)$$

Let $\lambda_{u1}, \ldots, \lambda_{uI}$, $\lambda_{v1}, \ldots, \lambda_{vI}$, $\lambda_{z1}, \ldots, \lambda_{zI}$ be the eigenvalues of the covariance functions of u, v, z (same eigenvalues of the covariance matrixes of U, V, U+V) respectively. From well known formula for the likelihood ratio in the white Gaussian channel (e.g. [6, Eq. (51)], [7, Eq. 4.8]), we have for the last three expressions in (24):

$$-D(\sqrt{q}\, \tilde{z} + w \| w) + \cos^2 \alpha\, D(\sqrt{q}\, \tilde{u} + w_1 \| w_1)$$
$$+ \sin^2 \alpha\, D(\sqrt{q}\, \tilde{v} + w_2 \| w_2)$$
$$= \frac{1}{2} \sum_{i=1}^{N} \ln(1 + q\lambda_{zi}) - \frac{1}{2} \sum_{i=1}^{N} q\lambda_{zi} - \frac{\cos^2 \alpha}{2} \sum_{i=1}^{N} \ln(1 + q\lambda_{ui})$$
$$+ \frac{\cos^2 \alpha}{2} \sum_{i=1}^{N} q\lambda_{ui} - \frac{\sin^2 \alpha}{2} \sum_{i=1}^{N} \ln(1 + q\lambda_{vi})$$
$$+ \frac{\sin^2 \alpha}{2} \sum_{i=1}^{N} q\lambda_{vi} = \frac{1}{2} \sum_{i=1}^{N} \ln \frac{1 + q\lambda_{zi}}{(1 + q\lambda_{ui})^{\cos^2 \alpha} (1 + q\lambda_{vi})^{\sin^2 \alpha}}$$
$$\to \frac{1}{2} \sum_{i=1}^{N} \ln \frac{\lambda_{zi}}{\lambda_{ui}^{\cos^2 \alpha}\, \lambda_{vi}^{\sin^2 \alpha}} \qquad q \to \infty.$$
$$(26)$$

Then, using the relation $D(X \| \tilde{X}) = \frac{N}{2} \sum_{i=1}^{N} \ln(2\pi e \lambda_{Xi}) - h(X)$

we have from (24), (25) and (26) for $q \to \infty$

$$h(U \cos \alpha + V \sin \alpha) \geq \cos^2 \alpha\, h(U) + \sin^2 \alpha\, h(V). \qquad (27)$$

Inequality (27) is equivalent to (13) – the Shannon EPI for random vectors (see e.g. [1]).

**Jacob Binia** was born in Jerusalem, Israel, on June 16, 1941. He received his B.Sc., M.Sc., and D.Sc. degrees in electrical engineering from the Technion-Israel Institute of Technology, Haifa, in 1963, 1968, and 1973, respectively. He joined the Armament Development Authority, Ministry of Defense, Israel, in 1963. From 1973 to 1990 he served first as Head of Signal Processing Group and then as Chief Engineer in the Communications Department, RAFAEL – Electronic Division. From 1992 to 1995 he was part of the National Electronic Warfare Research and Simulation Center.

During the years 1985, 1991, while on Sabbaticals from RAFAEL, he was invited by the Electrical Engineering Faculty at the Technion-Israel Institute of Technology to be a Guest Associate Professor.

He is now with New Elective - Engineering Services Ltd., Israel, where he is employed as a communications systems consultant.